\documentstyle[12pt,epsfig]{article}
\newcommand{\institute}[1]{\parbox{16cm}{%
\centering\normalsize \sl #1}}

\vspace{6cm}

\title{\bf Endpoint of the hot electroweak phase transition}
\author{%
F.~Csikor\\\\
\institute{Institute for Theoretical Physics, E\"otv\"os University,\\
H-1088 Budapest, Hungary}\\\\
Z.~Fodor%
\thanks{On leave from Institute for Theoretical Physics,
E\"otv\"os University, H-1088 Budapest, Hungary}\\\\
\institute{KEK, Theory Group, 1-1 Oho, Tsukuba 305, Japan}\\\\
J.~Heitger%
\thanks{Present address: DESY, Platanenallee 6, D-15738 Zeuthen, Germany}\\\\
\institute{Institut f\"ur Theoretische Physik I, Universit\"at M\"unster,\\
D-48149 M\"unster, Germany
}}

\date{}
\begin{document}
\maketitle

\vspace{-12.0cm}

{ \normalsize
\hfill \parbox[t]{4cm}{ITP-Budapest 541 \\ KEK-TH-580 \\MS-TPI-98-16\\
                  } \\[7em]
\vspace{6.2cm}

\begin{abstract} \noindent

We give the nonperturbative phase diagram of the four-dimensional 
hot electroweak phase transition.
The Monte-Carlo analysis is done on lattices with different
lattice spacings ($a$). A systematic extrapolation  $a \rightarrow 0$
is done. Our results show that the finite temperature SU(2)-Higgs 
phase transition is of first order for Higgs-boson masses 
$m_H<66.5 \pm 1.4$ GeV.
At this endpoint the phase transition is of second order, 
whereas above it only a 
rapid cross-over can be seen. The full four-dimensional result
agrees completely with that of the dimensional reduction 
approximation.  This fact is of particular importance, because
it indicates that the fermionic sector of the Standard Model  
can be included perturbatively.  We obtain that the Higgs-boson endpoint 
mass in the Standard Model is  $72.4 \pm 1.7$ GeV. Taking into account 
the LEP Higgs-boson mass lower bound excludes 
any electroweak phase transition in the Standard Model.

\end{abstract}

The observed baryon asymmetry is finally determined at the 
electroweak phase transition (EWPT) \cite{KuRS}.
The understanding of this asymmetry needs a quantitative
description of this phase transition. Unfortunately, 
the perturbative approach breaks down for the physically
allowed Higgs-boson masses (e.g. $m_H>70$ GeV) \cite{pert}.
In order to understand this nonperturbative phenomenon a 
systematically controllable technique is used, namely lattice
Monte-Carlo (MC) simulations. Since merely the bosonic sector is
responsible for the bad perturbative features (due to infrared problems)
the simulations are done without the inclusion of fermions. 
The first results dedicated to this questions were obtained
on four-dimensional lattices \cite{4d}. Soon after, simulations of the 
reduced model in three-dimensions were initiated, as another
approach \cite{3d}. This technique contains
two steps. The first is a perturbative reduction of the
original four-dimensional model to a three-dimensional one
by integrating out the heavy degrees of freedom. The
second step is the nonperturbative analysis of the
three-dimensional model on the lattice, which is less
CPU-time consuming than the MC simulation in 
the four-dimensional model. The comparison of the results
obtained by the two techniques is not only a useful 
cross-check on the perturbative reduction procedure
but also a necessity. The reason for that is 
that the fermions, which behave as the heavy bosonic modes,  
must be included perturbatively, anyhow. 

In the recent years exhaustive studies have been carried out
both in the four-dimensional \cite{4d-rev} and in the 
three-dimensional \cite{3d-rev} sectors of the problem.   
These works determined several cosmologically important quantities
such as the critical temperature ($T_c$), interface tension
($\sigma$) and latent heat ($\Delta \epsilon$).

Previous works show that the strength of the first order
EWPT gets weaker as the mass of the Higgs-boson 
increases. Actually the line of the first order phase
transitions, separating the symmetric and broken phases 
on the $m_H-T_c$ plane has an endpoint, $m_{H,c}$. There
are several direct and indirect evidences for that. In
four dimension at $m_H \approx 80$ GeV the EWPT turned out
to be extremely weak, even consistent with the no phase
transition scenario on the 1.5-$\sigma$ level \cite{4d80}.
Three-dimensional results show that for $m_H>95$ GeV
no first order phase transition exists \cite{3d95} and more specifically
that the endpoint is  $m_{H,c} \approx 67$ GeV \cite{3d80}. In this 
letter we present the analysis of the endpoint on four 
dimensional lattices. We study the thermodynamical limit of
the first Lee-Yang zeros of the partition function \cite{3d80}.
In order to get rid of the finite lattice spacing effects
a careful extrapolation to the continuum limit is performed. 
The endpoint value of the SU(2)-Higgs model is perturbatively transformed 
to the full Standard Model (SM).

We will study the four-dimensional SU(2)-Higgs lattice model 
on asymmetric lattices, i.e. lattices
with different spacings in temporal ($a_t$) and spatial
($a_s$) directions. Equal lattice spacings are used in the three
spatial directions ($a_i=a_s,\ i=1,2,3$) and another one 
in the temporal direction ($a_4=a_t$). The asymmetry of the
lattice spacings is given by the asymmetry factor $\xi=a_s/a_t$.
The different lattice spacings can be ensured by
different coupling strengths in the action for time-like and space-like
directions. The action reads  
\begin{eqnarray}\label{lattice_action}
&& S[U,\varphi]= \nonumber \\
&& \beta_s \sum_{sp}
\left( 1 - {1 \over 2} {\rm Tr\,} U_{pl} \right)
+\beta_t \sum_{tp}
\left( 1 - {1 \over 2} {\rm Tr\,} U_{pl} \right)
\nonumber \\
&&+ \sum_x \left\{ {1 \over 2}{\rm Tr\,}(\varphi_x^+\varphi_x)+
\lambda \left[ {1 \over 2}{\rm Tr\,}(\varphi_x^+\varphi_x) - 1 \right]^2
\right. \nonumber \\
&&\left.
-\kappa_s\sum_{\mu=1}^3
{\rm Tr\,}(\varphi^+_{x+\hat{\mu}}U_{x,\mu}\,\varphi_x)
-\kappa_t {\rm Tr\,}(\varphi^+_{x+\hat{4}}U_{x,4}\,\varphi_x)\right\},
\end{eqnarray}
where $U_{x,\mu}$ denotes the SU(2) gauge link variable,  $U_{sp}$ and
$U_{tp}$
the path-ordered product of the four $U_{x,\mu}$ around a
space-space or space-time plaquette, respectively;
$\varphi_x$ stands for the Higgs field. It is
useful to introduce the hopping parameter
$\kappa^2=\kappa_s\kappa_t$ and
$\beta^2=\beta_s\beta_t$. The anisotropies
$\gamma_\beta^2=\beta_t/\beta_s$ and $\gamma_\kappa^2=\kappa_t/\kappa_s$
are functions of the asymmetry $\xi$. These functions have been 
determined perturbatively \cite{T0pert} and non-perturbatively
\cite{T0nonpert} demanding the restoration of the rotational
symmetry in different channels. In this paper we use the asymmetry parameter
$\xi=4.052$, which gives $\gamma_\kappa=4$ and $\gamma_\beta=3.919$.
Details of the simulation techniques can be found in \cite{4d-rev}.

\begin{figure}\begin{center}
\epsfig{file=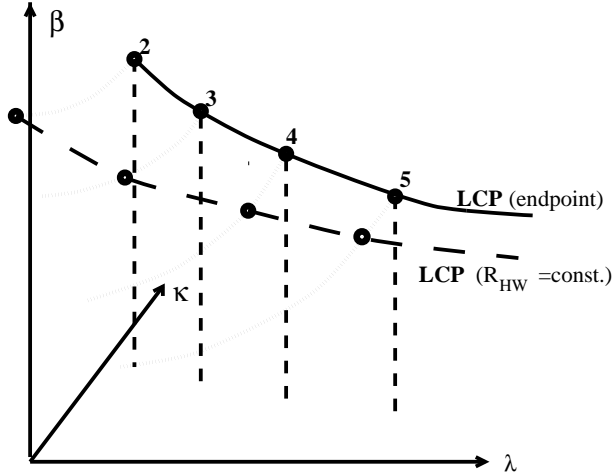,width=8.0cm}
\caption{\label{fig1}
{ \sl  Schematic view of the phase diagram. The solid line represents the
LCP defined by the endpoint condition. The numbers on the line correspond to
the temporal extension for which the endpoint is realized (the dashed
lines show  their projection to the $\kappa$ - $\lambda$ plane). The dotted
lines running into these numbered points correspond to first order phase 
transitions for $g_R^2$ = const. but different $R_{HW}$-s. A LCP defined 
by a constant $R_{HW}$ value is shown by the long dashed line.
}}
\end{center}\end{figure}

We have performed our simulations on finer and finer lattices,
moving along the lines of constant physics (LCP). In our case 
there are three bare parameters ($\kappa, \beta, \lambda $). The bare 
parameters are chosen in a way that the zero temperature renormalized gauge 
coupling $g_R$ is held constant and the mass ratio for the Higgs- and W-bosons 
$R_{HW}=m_H/m_W$ corresponds to the Higgs mass at the endpoint of 
first order phase transitions: $R_{HW,c}$. These two conditions determine 
a LCP as 
a one-dimensional subspace in the original space of bare parameters. 
The position on the LCP gives the lattice spacing $a$. As the lattice 
spacing decreases $R_{HW,c} \rightarrow R_{HW,cont.}$. A schematic 
illustration is shown in Fig. 1. The LPC (solid line) defined by the 
endpoint represents the above idea. The short dashed lines give the projections to the $\lambda - \kappa $ plane. The increasing numbers on the LCP show 
the temporal extensions of the lattice, thus corresponding to smaller and 
smaller lattice spacings. The dotted lines represent phase transition points of 
theories with fixed renormalized $g^2$ and $L_t$ but different $R_{HW}$ 
values. Along the dotted lines one can observe first order phase transitions 
upto the LCP defined by the endpoint condition. Note, however, that 
this endpoint LCP is not the same as the LCP defined by the constant 
$R_{HW} =R_{HW,cont.} $ value (long dashed line). They merge for 
decreasing lattice spacings, but at larger $a$ the difference is the result 
of the 'poor realization' of Wilson's RG transformations with only three terms 
and parameters in the action. It is worth mentioning that the SU(2)-Higgs model is trivial for small gauge couplings, therefore, the $a \rightarrow 0$ limit 
can not be performed. Even the points on the endpoint LCP do not define 
continuum theories. The second order phase transitions on it merely reflect a 
finite temperature phenomenon, the corresponding zero temperature SU(2)-Higgs 
theory is still trivial.

Since our theory is a bosonic one we assumed that the finite size corrections
are quadratic in the lattice spacings; therefore an $a^2 $ fit has been
performed for  $R_{HW,c} $ in order to determine its continuum value.

The technical implementation of the above LCP idea has been done as follows. 
By fixing $\beta=8.0$ in the simulations, we have observed that  $g_R $ is 
essentially constant within our errors.  For the small differences in $g_R $ 
we have performed perturbative corrections. We have carried out $T \neq 0$ 
simulations  on $L_t= 2,3,4,5$ lattices (for the finite temperature case 
one uses $L_t \ll L_x ,L_y ,L_z )$, and tuned $\kappa $ to the transition 
point.  This condition fixes the lattice spacings: $a_t =a_s / \xi \, = \, 
1/(T_c L_t ) $ in terms of the transition temperature $T_c $ in physical units.
The third parameter $\lambda $, finally specifying the physical Higgs mass 
in lattice units, has been chosen   in a way that the transition 
corresponds to the endpoint of the first order phase transition subspace. 

In this paper $V=L_t\cdot L_s^3$ type four-dimensional 
lattices are used. For each $L_t$ we had 8 different lattices, each
of them had approximately twice as large lattice-volume as the previous one.
The smallest lattice was $V=2\cdot 5^3$ and the largest one
was $V=5\cdot 50^3$. We collected quite a large statistics and
the Ferrenberg-Swendsen reweighting \cite{FS} was used to obtain information 
in the vicinity of a simulation point. 

The determination of the endpoint of the finite temperature
EWPT, thus a characteristic feature of the phase diagram,
is done by the use of the Lee-Yang zeros of the  
partition function ${\cal Z}$ \cite{LY}. 
Near the first order phase transition point the partition function reads

\begin{eqnarray}
{\cal Z}={\cal Z}_s + {\cal Z}_b \propto \exp (-V f_s ) + \exp ( -V f_b ) \, ,
\end{eqnarray}
where the indices s(b) refer to the symmetric (broken) phase and $f$ stands 
for the free-energy densities. Near the phase transition point we also have 
\begin{eqnarray}
f_b = f_s + \alpha (\kappa - \kappa _c ) \, ,
\end{eqnarray}

\begin{table*}
\begin{center}
\begin{tabular}{|c|c|c|c|c|}
\hline
$L_t$ &$\lambda_{sim.}$&$\kappa_{sim.}$ &$R_{HW}$&$g^2$ \\ \hline
2 &0.000178 & 0.107733 &0.934(10) & 0.569(4) \\ \hline
3 &0.000178 & 0.106988 &0.913(12) & 0.575(3) \\ \hline
4 &0.000178 & 0.106620 &0.905(8)  & 0.585(5) \\ \hline
5 &0.000178 & 0.1064974&0.867(36) & 0.566(30) \\
\hline
\end{tabular}
\caption{ \sl Summary of simulation parameters and results
on $R_{HW}$ and $g_R^2$ at $T=0$.
}
\end{center}
\end{table*}
since the free-energy density is continuous. It follows that 
\begin{eqnarray}
{\cal Z}=2 \exp [ -V ( f_s +f_b )/2 ] \cosh [ -V \alpha (\kappa -\kappa_c )] \, \
\end{eqnarray}
 which shows that for complex $\kappa$ ${\cal Z}=$ vanishes at 
 \begin{eqnarray}
 {\rm Im} (\kappa )= \pi \cdot (n-1/2) / (V\alpha )
 \end{eqnarray}
 for integer $n$.  In case a first order phase transition is present, 
 these Lee-Yang 
 zeros move to the real axis as the volume goes to infinity. In case a 
 phase transition is absent the Lee-Yang
  zeros stay away from the real $\kappa $ axis. Thus the way the Lee-Yang
   zeros move in this limit is a good indicator for the presence or 
   absence of a first order phase transition \cite{LY}
 Denoting
$\kappa_0$ the lowest zero of ${\cal Z}$, i.e. the  position of the zero 
closest zero to the real axis, one expects in the vicinity of
the endpoint the scaling law 
${\rm Im}(\kappa_0)=c_1(L_t,\lambda)V^\nu+c_2(L_t,\lambda)$.   
In order to pin down the endpoint we are looking
for a $\lambda$ value for which $c_2$ vanish. 
In practice we analytically continue ${\cal Z}$ to complex 
values of $\kappa $ by reweighting the available data. 
Also small changes  in
$\lambda$ have been done by reweighting. 
As an example, the dependence of $c_2$ on  $\lambda$ for $L_t =3$ 
is shown in fig. 2. To determine the critical value of $\lambda$ 
i.e. the largest value, where $c_2=0$, we 
have performed  fits linear in $\lambda$ to the nonnegative $c_2$ values.

Having determined the endpoint
$\lambda_{crit.} (L_t)$ for each $L_t$ we calculate the $T=0$
quantities ($R_{HW},g_R^2$) on $V=(32L_t)\cdot (8L_t)\cdot (6L_t)^3$ lattices,
where $32L_t$ belongs to the temporal extension, and 
extrapolate to the continuum limit. All the $T=0$ 
simulations were performed 
at $\lambda=0.000178$ and an extrapolation to the $\lambda_{crit.} (L_t)$  
has been made. The parameters and results of the simulations
are collected in table 1, while table 2 shows the $R_{HW}$ values extrapolated 
to the $\lambda_{crit.} (L_t)$. Having established the correspondence 
between $\lambda_{crit.} (L_t)$ and $R_{HW}$, the $L_t$ dependence of the 
critical $R_{HW}$ is easily obtained. Fig. 3 shows the dependence of 
the endpoint $R_{HW}$ values on $1/L_t^2 $. A linear extrapolation 
in $1/L_t^2 $ yields the infinite volume (i.e. continuum limit) 
value of the endpoint 
$R_{HW}$. We obtain $66.5 \pm 1.4$ GeV, which is our final result.

\begin{table}[htb]
\begin{center}
\begin{tabular}{|c|c|c|c|}
\hline
$L_t$ &$\lambda_{crit.}$&$\kappa_{crit.}$&$R_{HW,c}$\\ \hline
2&0.0001773(14)&0.1077292(2) & 0.932(10)\\ \hline
3&0.0001664(27) &0.1069581(2) &  0.883(12)\\ \hline
4&0.0001590(44) &0.1066316(3) &  0.856(8)\\ \hline
5& 0.0001664(20) &0.1064948(6) & 0.838(36)\\
\hline
\end{tabular}
\caption{ \sl Critical $\lambda$ corresponding to the endpoint of phase
transition as function of $L_t$ and the corresponding value of $R_{HW}$.
}
\end{center}
\end{table}

Comparing out result to those of the 3d analyses \cite{3d80} one 
observes complete 
agreement. Since the error bars on the endpoint determinations are on the 
few percent level, the uncertainty of the dimensional reduction procedure 
is also in this range. This indicates that the analogous perturbative 
inclusion  of the fermionic sector results also in few percent error on 
$M_H$.

Based on our published data \cite{4d-rev,T0nonpert} and the results of 
this paper  we are now able to draw the precise phase diagram of the 
SU(2)-Higgs model in the ($T_c /m_H - R_{HW} $) plane. This is shown in 
fig. 4. The continuous line -- representing the phase-boundary -- is 
a quadratic fit to the data points.

\begin{figure}\begin{center}
\epsfig{file=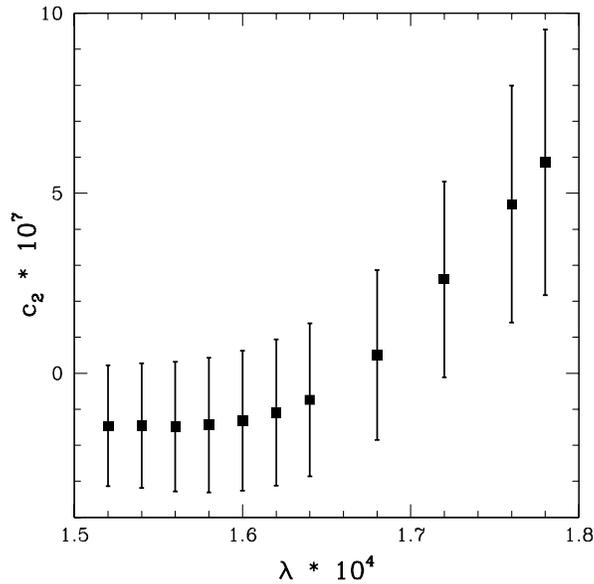,width=8.0cm}
\caption{\label{fig2}
{ \sl Dependence of $c_2$ on $\lambda$ for $L_t=3$.
}}
\end{center}\end{figure}

Finally, we determine what is the endpoint value in the full SM. 
Our nonperturbative analysis shows that the perturbative integration of 
the heavy modes is correct within our error bars. Therefore we use perturbation theory \cite{KLRS96} to transform the SU(2)-Higgs model endpoint value to 
the full SM. We obtain $72.4 \pm 1.7$ GeV, where the error 
includes the measured error of $R_{HW,cont.}$, $g_R^2$ and the estimated 
uncertainty \cite{Laine96} 
due to the different definitions of the 
gauge couplings between this paper and \cite{KLRS96}. The dominant error 
comes from the uncertainty on the position of the endpoint.

In conclusion, we have determined the endpoint of hot EWPT
 with the technique of Lee-Yang zeros from simulations 
in four-dimensional SU(2)-Higgs model. The phase diagram has been also 
presented. The phase transition is first order for Higgs masses less than 
$66.5 \pm 1.4$ GeV, while for larger Higgs masses only a rapid cross-over 
is expected. One of the most important results of the present letter is 
that integrating out the heavy modes  perturbatively is precise 
as shown by a comparison to our nonperturbative results. Thus the above 
$66.5 \pm 1.4$ GeV value can  be perturbatively  transformed  to
the full SM. We obtain $72.4 \pm 1.7$ GeV for the endpoint 
Higgs mass. As pointed out above the perturbative inclusion of the fermionic 
 sector of the SM is also correct to a few percent error level.  

 \begin{figure}\begin{center}
\epsfig{file=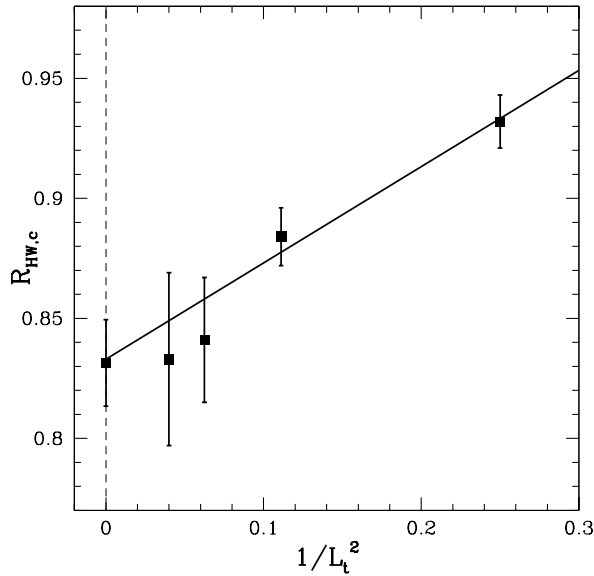,width=8.0cm}
 \caption{\label{fig3}
 { \sl Dependence of $R_{HW,c}$, i.e. $R_{HW}$ corresponding to the endpoint
 of first order phase transitions on $1/L_t^2$ and extrapolation to the
 infinite volume limit.
 }}
\end{center}\end{figure}

The present experimental lower limit of the SM Higgs-boson mass 
is $89.8$ GeV \cite{LEP}. Taking into account all errors (in particular 
those coming from integrating out the heavy fermionic modes), 
our endpoint value excludes the 
possibility of any EWPT in the SM. 
This also means that the SM baryogenesis in the early Universe 
is ruled out.

More details of this investigation will be published in a forthcoming 
publication \cite{tobe}.

We thank I. Montvay and R. Sommer for discussions.
Simulations have been carried out on the Cray-T90 at HLRZ-J\"ulich,
on the APE-Quadrics at DESY-Zeuthen and on the PMS-8G PC-farm
in Budapest. This work was partially supported by Hungarian Science Foundation
grants No. OTKA-T016240/T022929 and FKP-0128/1997.

\vspace{1cm}

\begin{figure}\begin{center}
\epsfig{file=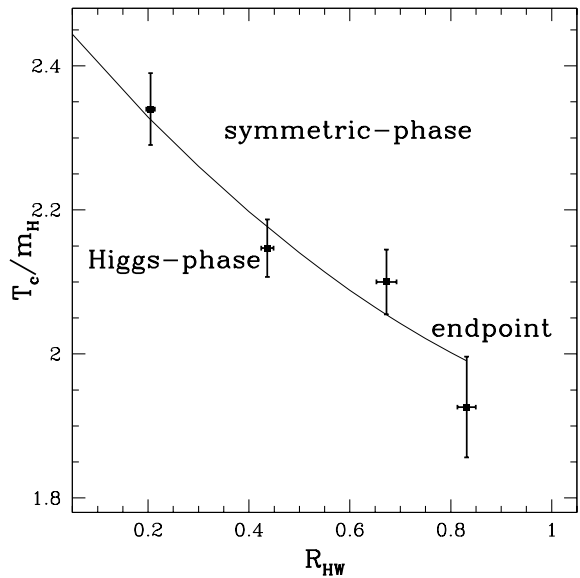,width=8.0cm}
\caption{\label{fig3v}
{ \sl  Phase diagram of the SU(2)-Higgs model in the ($T_c /m_H - R_{HW} $)
plane. The continuous line -- representing the phase-boundary -- is a quadratic fit
 to the data points.
 }}
\end{center}\end{figure}

\end{document}